\RequirePackage{lineno}
\documentclass[aps,prd,onecolumn,superscriptaddress,11pt]{revtex4}
\usepackage{graphicx}
\usepackage{amsmath,amssymb}
\usepackage{color}

\usepackage{mathrsfs}
\usepackage[x11names]{xcolor}
\usepackage{aas_macros}
\usepackage[normalem]{ulem}

\begin{document}
%\linenumbers

%%%%%%%%%%%%%%%%%%%%%%%%%%%%%%%%%%%%%%%%%%%%%%%%%%
%%%%%%%%%%%%%%%%%%%%%%%%%%%%%%%%%%%%%%%%%%%%%%%%%%

\title{Linking high-energy cosmic particles by black-hole jets\\
embedded in large-scale structures}
\author{Ke Fang}
\affiliation{Department of Astronomy; Joint Space-Science Institute, University of Maryland, College Park, MD, 20742-2421, USA}
\author{Kohta Murase}
\affiliation{Department of Physics, Pennsylvania State University, University Park, PA 16802, USA}
\affiliation{Department of Astronomy and Astrophysics, Pennsylvania State University, University Park, PA 16802, USA}
\affiliation{Center for Particle and Gravitational Astrophysics, Pennsylvania State University, University Park, PA 16802, USA}
\affiliation{Yukawa Institute for Theoretical Physics, Kyoto University, Kyoto 606-8502, Japan}

\date{published online in Nature Physics Letters on 22 January 2018; accepted on 23 November 2017}

\begin{abstract}
The origin of ultrahigh-energy cosmic rays (UHECRs) is a half-century old enigma~\cite{PhysRevLett.10.146}. The mystery has been deepened by an intriguing coincidence: over ten orders of magnitude in energy, the energy generation rates of UHECRs, PeV neutrinos, and isotropic sub-TeV $\gamma$ rays are comparable, which hints at a grand-unified picture~\cite{2016PhRvD..94j3006M}. Here we report that powerful black hole jets in aggregates of galaxies can supply the common origin of all of these phenomena. Once accelerated by a jet, low-energy cosmic rays confined in the radio lobe are adiabatically cooled; higher-energy cosmic rays leaving the source interact with the magnetized cluster environment and produce neutrinos and $\gamma$ rays; the highest-energy particles escape from the host cluster and contribute to the observed cosmic rays above 100~PeV. The model is consistent with the spectrum, composition, and isotropy of the observed UHECRs, and also explains the IceCube neutrinos and the non-blazar component of the Fermi $\gamma$-ray background, assuming a reasonable energy output from black hole jets in clusters. 
\end{abstract}

\pacs{95.85.Ry, 98.70.Sa, 98.70.Vc\vspace{-0.3cm}}
% 95.85.Ry Neutrino, muon, pion, and other elementary particles; cosmic rays
\maketitle
%%%%%%%%%%%%%%%%%%%%%%%%%%%%%%%%%%%%%%%%%%%%%%%%%%
%%%%%%%%%%%%%%%%%%%%%%%%%%%%%%%%%%%%%%%%%%%%%%%%%%

% UHECR
The origin of ultrahigh-energy cosmic rays (UHECRs) is still unknown~\cite{Hillas84}. Measurements by the Pierre Auger Observatory (Auger)~\cite{Aab:2015bza} and the Telescope Array (TA)~\cite{TA_ICRC15} find a power-law spectrum, $\Phi\propto E^{-2.6}-E^{-2.7}$ (where $E$ is particle energy and $\Phi$ is the diffuse intensity in units of particles per energy, area, time, and solid angle), with a decline above $6\times{10}^{19}\,\rm eV$, probably due to the interaction of UHECRs with cosmic radiation backgrounds such as the cosmic microwave background (CMB) or an upper limit of the particle energy reachable by the accelerator. Small-scale anisotropy in their arrival directions has not been established~\cite{Aab:2015bza,TA_ICRC15}.

% neutrino 
The IceCube Observatory recently discovered high-energy cosmic neutrinos~\cite{Halzen:2016gng,Aartsen:2013bka}, which have been anticipated to provide crucial clues to this age-old mystery. An astrophysical flux in the $0.1-1$~PeV range is found at the level of $10^{-8} \,\rm GeV~cm^{-2}~s^{-1}~sr^{-1}$ per flavor~\cite{Halzen:2016gng,Aartsen:2013bka,Aartsen:2016xlq, Aartsen:2016xlq2}, which is consistent with expectations of cosmic-ray ``reservoir'' models~\cite{Murase:2008yt,Kotera:2009ms,Loeb:2006tw,Murase:2013rfa}.  
The arrival directions of the observed events present no significant clustering, and indicate that the sources are extragalactic~\cite{Halzen:2016gng,Aartsen:2016xlq, Aartsen:2016xlq2}. 

% extragalactic gamma-ray 
A $\gamma$-ray counterpart is expected from the hadronic processes responsible for neutrino production.  If the source environment is transparent to $\gamma$ rays, these side products should show up at $1-100$~GeV energies after cascading in the extragalactic background light (EBL). A significant fraction of the extragalactic $\gamma$-ray background (EGB)~\cite{FermiBG,PhysRevLett.116.151105} measured by the Fermi Gamma-Ray Space Telescope may be explained by neutrino sources~\cite{Murase:2013rfa}. 
Despite the unknown origins of these multi-messenger emissions and fine structures in their data (such as a possible excess in the $10-100$~TeV neutrino spectrum~\cite{Halzen:2016gng}), it is remarkable that over ten orders of magnitude in energy, the energy generation rates of UHECRs, IceCube neutrinos, and Fermi EGB are all comparable~\cite{Murase:2013rfa,2016PhRvD..94j3006M}. 

% What is new?
Recent UHECR observations have revealed additional characteristic features of extragalactic cosmic rays. First, a hardening in the spectrum of light particles is seen around 100~PeV~\cite{::2013dga,Buitink:2016nkf}, right in the energy range where a steepening in the spectrum of heavy primary particles is observed (which is often called the ``second knee"). 
Second, a transition from light elements to medium-to-heavy elements around ${10}^{19}$~eV is suggested by the Auger data~\cite{Aab:2015bza}, and a heavy UHECR composition is also supported by the non-detection of cosmogenic neutrinos~\cite{2016arXiv160705886I}. Although the interpretation of the UHECR composition is still debated, direct measurements of indicators of the particle mass seem to be consistent between different experiments.
These features were not considered in the simplest convergence theory~\cite{2016PhRvD..94j3006M}; here we provide a concrete astrophysical model in which black hole jets embedded in large-scale structures reconcile these observations.
 
Relativistic jets from the accretion onto supermassive black holes provide promising sites for UHECR acceleration. The Hillas condition suggests that an active galactic nucleus (AGN) can accelerate a particle with charge $Z$ to a maximum energy, $E_{\rm max}\sim Z\,10^{19}\,\rm eV$, in jets or at external shocks that are known to be sites of leptonic emissions~\cite{Hillas84,2012ApJ...749...63M}. 
The energy spectrum of particles accelerated by the Fermi mechanism can be described by a power law, $dN_{\rm acc}/dE\propto E^{-s_{\rm acc}}$, with an index $s_{\rm acc}\sim2-2.5$.
Radio observations often find extended lobes (sometimes referred as bubbles or cocoons, which are plasma cavities inflated by the jet), with $10-100$ kiloparsec scales~\cite{2007MNRAS.381.1548K} and $0.1-10$ microgauss-level magnetic fields~\cite{2005ApJ...622..797K}.  
Particles with energies below $E_{{\rm lobe},c} = Z\,e\,B_{\rm lobe}\,l_{{\rm lobe},c}\sim1.4\times{10}^{18}\,Z\,\left(B_{\rm lobe}/5\,\mu\rm G\right)\,\left(l_{{\rm lobe},c}/0.3\,\rm kpc\right)\,\rm eV$ have a Larmor radius $r_L =E/(Ze\,B)$ that is much smaller than the coherence length of their magnetic structure, where $B_{\rm lobe}$ and $l_{{\rm lobe},c}\sim(0.01-0.1)\,r_{\rm lobe}$ are the magnetic field strength and the coherence length, and $r_{\rm lobe}$ is the lobe size and $c$ is the speed of light. Unlike relativistic electrons that cool inside jets or lobes, high-energy ions diffuse for $t_{\rm diff}^{\rm lobe} \sim6.1\,\left( {r_{\rm lobe}}/{10\,\rm kpc}\right)^2\, \left({E}/{Z\,1\,\rm PeV}\right)^{-1/3}\,\left({l_{{\rm lobe},c}}/{0.3\,\rm kpc}\right)^{-2/3}$ $\left({B_{\rm lobe}}/{5\,\mu \rm G}\right)^{1/3}\,{\rm Myr}$ and can enter the intracluster medium (ICM). Meanwhile, these high-energy ions suffer from adiabatic losses due to the expansion of the cocoon. The characteristic cooling time is $t_{\rm ad}\sim4.9\,\left(r_{\rm lobe}/10\,\rm kpc\right)$ $\left(v_{\rm lobe}/2000\,\rm km\,s^{-1}\right)^{-1}\,\rm Myr$, where $v_{\rm lobe}$ is the typical expansion velocity at source ages of $\sim0.1-10$~Myr~\cite{2011MNRAS.412.1229B}. Particles with energies above $E_{{\rm lobe},c}$ are less impacted, escaping semi-diffusively with $t_{\rm diff}^{\rm lobe}\propto E^{-2}$. Considering the competition between diffusion and cooling, we approximate the spectrum of cosmic rays leaking into the cluster to be $dN_{\rm inj}/dE\propto E^{-s_{\rm acc}}\,\exp(-t_{\rm diff}^{\rm lobe}/t_{\rm ad})$.

Radio-loud AGN activity that leads to cosmic-ray injections would preferentially reside in the centers of rich clusters~\cite{2007MNRAS.379..894B}.  
A cluster with a halo mass of $M=10^{14}\,M_{14}\,M_\odot$ has a virial radius $r_{\rm vir}\sim 1.2\,M_{14}^{1/3}\,\rm Mpc$.  The distribution of thermal gas is often described using the $\beta$ model as $n_{\rm ICM}(r)\propto\left[1+\left({r}/{r_c}\right)^2\right]^{-3\beta/2}$, where $\beta\approx0.8$ and $r_c\sim0.1\,r_{\rm vir}$ is the core radius~\cite{2014IJMPD..2330007B}. 
Turbulent magnetic fields in the ICM, which are probably induced by accretion shocks and other cluster dynamics, typically have a strength of a few~$\mu$G in the cluster center~\cite{2014IJMPD..2330007B}.  Assuming flux conservation and that the field traces the baryon distribution, we adopt a magnetic field profile $B(r)=B_0\left[1+\left( {r}/{r_{\mathrm{c}}}\right)^2\right]^{-\beta}$ with $B_0\sim 5\mu G$. 

\begin{figure}
\centering\includegraphics[scale=0.65]{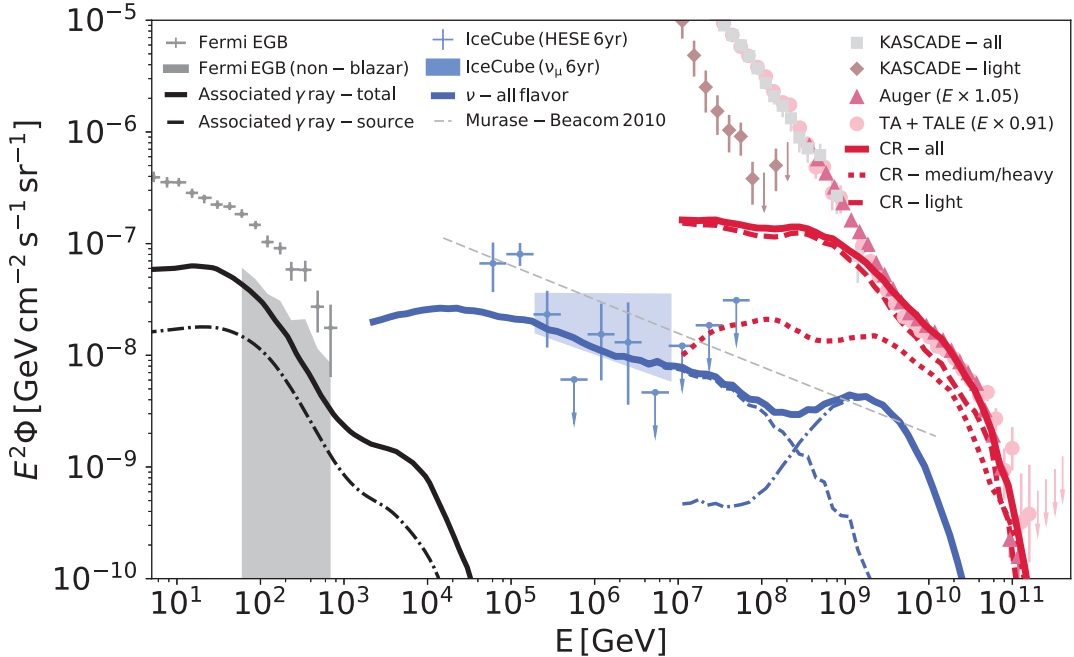}
\caption{\label{fig:spectrum} \small {\bf Extragalactic multi-messenger (UHECR, high-energy neutrino and $\gamma$-ray) background spectra from galaxy clusters and groups with black hole jets as cosmic-ray accelerators.} Measurements from the KASCADE-Grande~\cite{::2013dga}, Telescope Array and Telescope Array Low Energy extension (TALE)~\cite{TA_ICRC15}, Pierre Auger Observatory~\cite{Aab:2015bza} (with Auger energy scaled up by 5\% and TA energy scaled downed by 9\% to match the two measurements~\cite{2017arXiv170509111V}), IceCube~\cite{Aartsen:2016xlq, Aartsen:2016xlq2}, and Fermi Gamma-Ray Space Telescope~\cite{FermiBG,PhysRevLett.116.151105}. The total cosmic-ray spectrum (solid red) is decomposed into two composition groups: light (dashed red; H and He) and medium-heavy (dotted red; CNO, Si, Fe).  
PeV neutrinos (solid blue) are produced by interactions between cosmic rays and the ICM (dashed blue), and by UHECRs interacting with the CMB and EBL during their intergalactic propagation (dash-dotted blue).  The upper bound on the neutrino flux of UHECR nuclei (for $s_{\rm acc}=2.3$) is shown for reference (dashed grey)~\cite{2010PhRvD..81l3001M}. 
The $\gamma$-ray counterparts (solid black for the total flux and dash-dotted black for $\gamma$ rays produced in the ICM) are comparable to the non-blazar component of the EGB measured by the Fermi Gamma-Ray Space Telescope \cite{PhysRevLett.116.151105}.}
\end{figure}

Cosmic rays leaving the acceleration site and lobe enter the ICM of the host cluster (which functions as a cosmic-ray reservoir~\cite{Murase:2008yt,Kotera:2009ms}). The highest-energy ions travel in a straight line through the ICM. Particles reaching an energy $E_{c}\sim2\times10^{19}\,Z\,B_{-6}\,(l_{c}/{20\,\rm kpc})\,\rm eV$ have a gyro-radius comparable to the typical scales of magnetic field fluctuations in massive clusters, with $l_{c}$ about 1-10\% of the virial radius~\cite{2014IJMPD..2330007B}. Ions with energies well below $E_{c}$ propagate diffusively in the turbulent magnetic field of the cluster. The confinement, which could last for $\sim1-10$~Gyr depending on the particle energy, leads to efficient interactions of cosmic-ray nuclei with baryons and infrared background photons in the cluster, producing pions that decay into neutrinos and $\gamma$ rays via $\pi^{\pm}\rightarrow\nu_e(\bar{\nu}_e)+e^\pm+\nu_\mu+\bar{\nu}_\mu$ and $\pi^0\rightarrow2\gamma$, respectively.  
Finally, particles that leave the cluster propagate to the Earth through the intergalactic medium and extragalactic magnetic fields. UHECRs from sources beyond the energy-loss horizon are depleted via photodisintegration, photomeson production and Bethe-Heitler pair production processes with the CMB and the EBL, producing cosmogenic neutrinos peaked around EeV and $\gamma$ rays that cascade down to GeV-TeV energies.
 
We numerically simulate the propagation of cosmic rays in the magnetized ICM and from the source to the observer. We assume that a jetted source as a cosmic-ray accelerator can be anywhere in the core of a cluster with equal probability. We inject five representative groups of elements: hydrogen ($^1$H), helium ($^4$He), nitrogen ($^{14}$N), silicon ($^{28}$Si) and iron ($^{56}$Fe) according to the abundances of elements in Galactic cosmic rays (see Supplementary Information for details), and let each group follow the same power-law spectrum with a cutoff above the maximum rigidity, $dN_{\rm inj}/dR\propto R^{-s_{\rm acc}}\,\exp(-R/R_{\rm max})$, where $R=E/Ze$ is the rigidity, $s_{\rm acc}=2.3$, and $R_{\rm max}=2\times10^{21}/26\,{\rm V}$. We assume that ions are confined up to $t_{\rm inj}= 2\,\rm Gyr$, given that the peak period of AGN activity effectively lasts for $\sim2-3$~Gyr (see Supplementary Information for discussions on model uncertainties and details). 
The redshift evolution of the source density is taken to be $F(z)=(1+z)^{3}$ up to $z_{c}=1.5$, but its moderate variations barely impact our results. The cumulative flux is obtained by~\cite{Murase:2008yt}:
\begin{equation}
\Phi(E)= \frac{1}{4\pi} \int\frac{c\,dz}{H(z)}\,F(z)\, \int_{M_{\rm min}}^{\infty}\,dM\,\frac{dn}{dM}\,\frac{d\dot{N}}{dE'}(M,z),
\end{equation}
where $n$ is the halo number density, $dn/dM$ is the halo mass function, $H(z)$ is the Hubble parameter at redshift $z$, $d\dot{N}/dE'$ is the production rate of neutrinos (or propagated cosmic rays) from a given cluster with a redshifted energy $E'=(1+z)\,E$. 
We consider clusters with a halo mass above $M_{\rm min} = 5\times10^{13}\,M_\odot$ (corresponding to $\sim10^{11}\,M_\odot$ for the stellar mass of the main halo), which present higher radio-loud AGN fractions~\cite{2007MNRAS.379..894B}. 
For the intergalactic propagation, we assume that cosmic rays from a galaxy cluster have 50\% chance of encountering magnetic structures with an average strength of 2~nG and a coherence length of 1~Mpc.

\begin{figure}
\centering\includegraphics[scale=0.60]{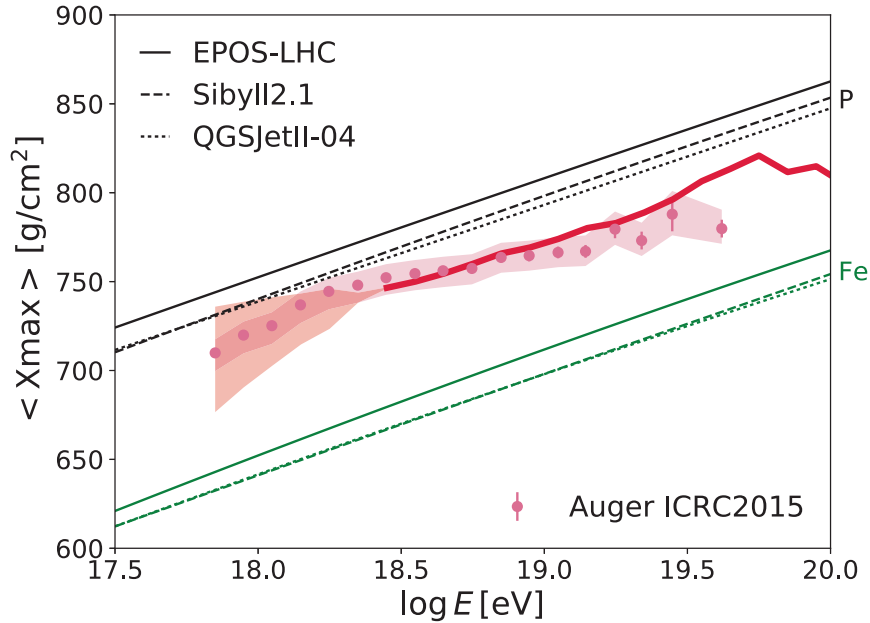}
\caption{\label{fig:Xmax}  \small {\bf Mean of the maximum depth of an air shower of UHECRs.} Values of $\langle X_{\rm max} \rangle$ for the UHECRs in Fig~\ref{fig:spectrum} (solid red line, calculated with the EPOS interaction model~\cite{2013JCAP...07..050D}) are compared with that of the Auger data~\cite{Aab:2015bza} (pink data points with the shaded region indicating systematic errors).  
For reference, $\langle X_{\rm max} \rangle$ of a 100\% proton (black) and 100\% iron nuclei composition (green) are shown, computed using three interaction models, EPOS-LHC (solid), Sybill 2.1 (long dashed), and QGSJetII-04 (short dashed). The red shaded region indicates the energy range where the extragalactic contribution is less than 85\% of the measured flux, and is determined by assuming that the residual flux, which could be a Galactic component, has a composition between proton and iron.}
\end{figure}
 
Figure~\ref{fig:spectrum} shows the integrated spectra of UHECRs and neutrinos from overdense regions with black hole jets. 
The normalization of the spectra is determined by a combined fit to the Auger spectral and $\langle X_{\rm max}\rangle$ data above $10^{18.45}$~eV, and the IceCube data above $2\times{10}^{14}$~eV. The goodness fit results in a $\chi^2=44.5$ for 30 degrees of freedom, corresponding to a p-value of 0.043 for this fiducial case. The cosmic-ray confinement in the lobe and the host cluster makes the injection spectrum harder below the second knee~\cite{Murase:2008yt,Murase:2013rfa}.
The spectral shape is agreement with measurements by both Auger and TA above $10^{18}\,\rm eV$. 
Primary and secondary cosmic-ray particles received by the observer are divided into two composition groups: light (including H and He) and intermediate/heavy (including CNO, Si, Mg, Fe), with the two crossing around $10^{19.5}\,\rm eV$. The mean of the maximum depth of an air shower, $\langle X_{\rm max}\rangle$, which depends on the mass of the UHE nucleon or nucleus, is shown in Figure~\ref{fig:Xmax}. The trend follows the $\langle X_{\rm max}\rangle$ data measured by Auger. Below $10^{18}\,\rm eV$, accounting for a Galactic contribution with $\Phi\propto E^{-3.4}$, the predicted cosmic-ray spectrum matches the light component of the KASCADE-Grande data~\cite{::2013dga}.

The neutrino spectrum is composed of two parts. 
Between $10^{14}\,\rm eV$ and $10^{17}\,\rm eV$, it is mostly contributed by particle interactions in the ICM. It agrees with the IceCube measurements above $10^{14}\,\rm eV$.  The low-energy neutrino spectrum is harder than that of accelerated cosmic rays, and the spectral steepening above $10^{15}\,\rm eV$ results from the faster escape of higher-energy cosmic rays.  Above $10^{18}\,\rm eV$, the neutrino flux is dominated by the cosmogenic neutrinos produced when UHECRs interact with the CMB and the EBL, and is consistent with the IceCube constraints at extremely high energies~\cite{2016arXiv160705886I}.  
Likewise, the observed sub-TeV $\gamma$ rays are produced {\it both} in the ICM and during intergalactic propagation~\cite{2016PhRvD..94j3006M}. Thanks to the hard injection spectrum, the total $\gamma$-ray flux largely originates from electromagnetic cascades, and is consistent with the non-blazar component of the EGB~\cite{PhysRevLett.116.151105}. In addition to the hard $\gamma$-ray spectrum, our model also predicts a dominance of low-mass clusters, and the $\gamma$-ray and radio limits from individual clusters~\cite{2015A&A...578A..32Z} can be satisfied. 
 
% energetics 
The chance of previously or currently having active jets in a cluster, $f_{\rm jet}$, and the average cosmic-ray luminosity of contained active galaxies per cluster, $L_{\rm CR}$, are left as free parameters. Assuming $L_{\rm CR}\sim10^{44}-10^{45}\,\rm erg\,s^{-1}$, we obtain $f_{\rm jet}\sim10-100\%$. This is consistent with duty cycles of the accretion-driven evolution of black holes~\cite{2011ApJ...740...51M}. The number density of clusters and groups with a mass above $5\times10^{13}\,M_\odot$ is $\sim{\rm a~few}\times10^{-5}\,\rm Mpc^{-3}$. This satisfies the muon neutrino limits on the neutrino source density derived from the absence of multiplets~\cite{2016PhRvD..94j3006M}, as well as the lower bounds on the UHECR source density derived from the lack of strong anisotropy in the UHECR data~\cite{AugerBound}. 
Our model predicts an association between directions of neutrino events and low-mass clusters that have past and ongoing jet activities of supermassive black holes. Alongside multiplet signals from nearby candidate sources, this correlation can be tested~\cite{2016PhRvD..94j3006M} with future neutrino observations by experiments such as IceCube-Gen2.
%%%%%%%%%%%%%%%%%%%%%%%%%%%%%%%%%%%%%%%%%%%%%%%%%%
%%%%%%%%%%%%%%%%%%%%%%%%%%%%%%%%%%%%%%%%%%%%%%%%%%

\medskip
{\bf Data availability}\\
The authors declare that the data supporting the plots within this paper and other findings of this study are available from the authors upon reasonable request. The data of Figures 1 and 2 of the main text can be found at https://figshare.com/s/8216c3831633e29dace3.

%\acknowledgements
{\bf Acknowledgements}\\
We thank Rafael Alves Batista, Mauricio Bustamante, M. Coleman Miller, Chris Reynolds and Michael Unger for helpful comments. 
This work made use of supercomputing resources at the University of Maryland. We gratefully acknowledge support from the Eberly College of Science of Penn State University and the Institute for Gravitation and the Cosmos. The work of K.M. is supported by Alfred P. Sloan Foundation and NSF grant No. PHY-1620777. 

{\bf Author contributions}\\
K.F. performed simulations and produced the figures. K.M. designed the research and contributed to the calculations. Both authors equally edited the manuscript.

{\bf Competing financial interests}\\
The authors declare no competing financial interests.

{\bf Additional information}\\
Reprints and permissions information is available at www.nature.com/reprints. Correspondence and requests for materials should be addressed to K.M.

{\bf Note added}\\
After this Letter was submitted to Nature Physics, the preprint arXiv:1704.06893 appeared with an incorrect quote about our work. Our reservoir model indeed explains diffuse spectra of neutrinos, $\gamma$ rays and UHECRs ``dominantly'', as well as the UHECR composition and sub-ankle cosmic rays.

%%%%%%%%%%%%%%%%%%%%%%%%%%%%%%%%%%%%%%%%%%%%%%%%%%
%%%%%%%%%%%%%%%%%%%%%%%%%%%%%%%%%%%%%%%%%%%%%%%%%%

%\input{ms.bbl}
\bibliography{FM17arXiv.bib}
%\begin{thebibliography}{99}
%\end{thebibliography}

\clearpage

%%%%%%%%%% Merge with supplemental materials %%%%%%%%%%
\setcounter{equation}{0}
\setcounter{figure}{0}
\setcounter{table}{0}
\setcounter{section}{0}
\setcounter{page}{1}
\makeatletter
\renewcommand{\theequation}{S\arabic{equation}}
\renewcommand{\thefigure}{S\arabic{figure}}
\renewcommand{\thetable}{S\arabic{table}}
\newcommand\ptwiddle[1]{\mathord{\mathop{#1}\limits^{\scriptscriptstyle(\sim)}}}

\appendix

\section*{\Large Supplementary Information}

\section*{Methods}
Cosmic rays can be accelerated by black hole jets by mechanisms such as shock acceleration, shear acceleration, and magnetic reconnection.  After escaping from the jet and lobe/bubble, cosmic rays rectilinearly or diffusively propagate in the turbulent magnetic field of the host cluster (or group). 
In our grand-unified model, while we consider AGN jets as the principal cosmic-ray accelerators, we treat galaxy clusters and groups as cosmic-ray reservoirs, in which PeV neutrinos and a fraction of sub-TeV $\gamma$ rays are produced~\cite{Murase:2008yt,Murase:2009zz,Kotera:2009ms}.
The semi-diffusive propagation is implemented based on the public UHECR propagation code CRPropa 3~\cite{2016JCAP...05..038A}. The diffusive propagation is computed semi-analytically by letting particles random walk with a step size equal to the coherence length of the magnetic field. The numerical step size is then converted to an actual trajectory length through the on-site diffusion coefficient~\cite{Kotera08a,FO16}. 
Interactions between nuclei and target nucleons of the ICM gas are computed using pre-tabulated cross sections and products that were calculated with the hadronic interaction model EPOS~\cite{PhysRevC.74.044902}. 
Photomeson productions (photodisintegration) due to interactions between cosmic-ray protons (nuclei) and the infrared background of clusters are computed based on SOPHIA~\cite{2000CoPhC.124..290M} (TALYS~\cite{TALYS}) through CRPropa 3. 

Galaxy clusters and groups, in which AGN and other cosmic-ray accelerators are embedded, provide promising sites for neutrino and gamma-ray production~\cite{Berezinsky:1996wx,Murase:2012rd}. The density profile of the intracluster medium gas in a cluster is normalized by $f_{\rm b}M/(\mu\,m_p)=\int n_{\rm ICM}(r)dV$, where $f_b$ is the gas fraction in galaxy clusters, which is compatible with the cosmic mean baryon fraction $\Omega_b/\Omega_m= 0.167$~\cite{2011ApJS..192...18K, 2013A&A...550A.131P}, $\mu\approx0.61$ is the mean molecular weight, and $m_p$ is the mass of a proton. The infrared background photons in the cluster is modeled following reference~\cite{Takami:2012gu} (scaled up by a factor of 3 to be consistent with reference~\cite{Kotera:2009ms}), with a spectral energy distribution resulting from the superposition of the emission of 100 giant elliptical galaxies, and a density profile following the ICM gas distribution. The infrared background of the cluster add to the CMB and the EBL as target radiation fields. 
While not only radio-loud AGN but also radio-quiet AGN have been suggested as UHECR accelerators~\cite{Peer:2009vnw}, the former objects are known to be more powerful. Radio-loud AGN are more common as a central galaxy in cool core clusters than in non-cool core clusters. To be conservative, we have not taken into account the high densities at the center of the subset of clusters that have cool cores~\cite{Kotera:2009ms}.  

\begin{figure*}[tb]
\includegraphics[width=1.0\linewidth]{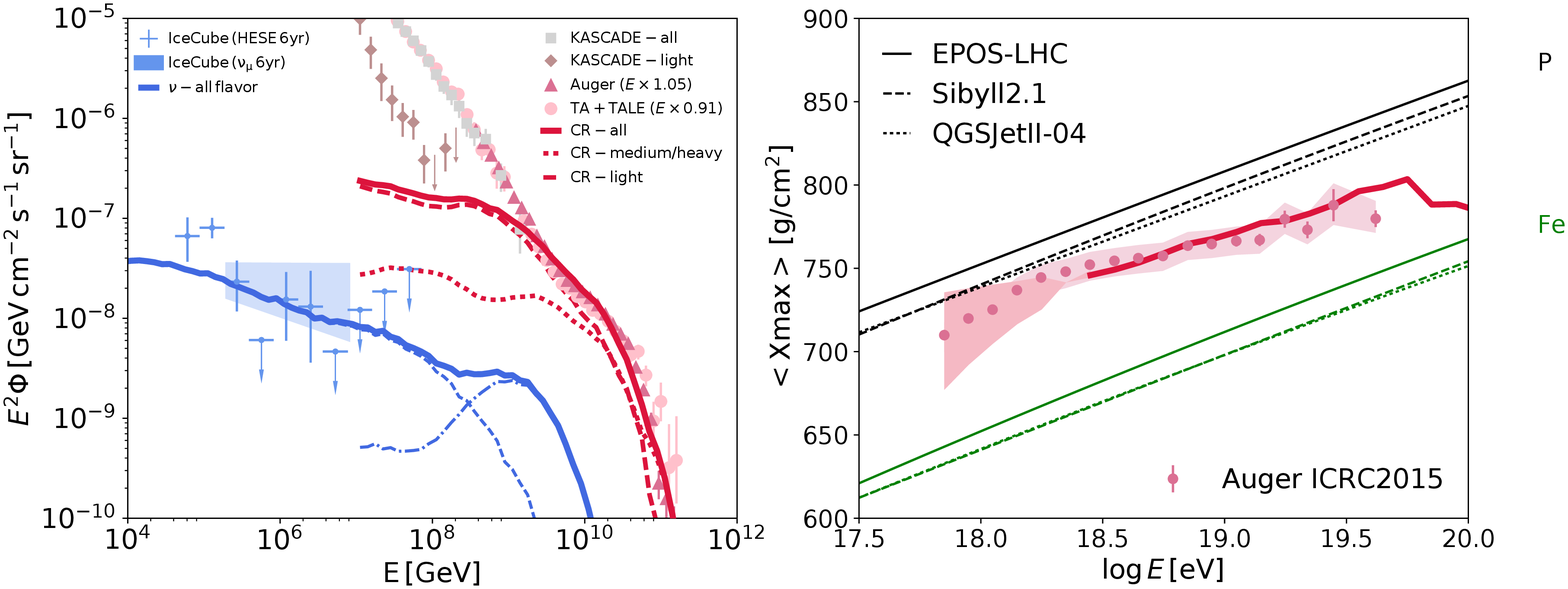}
\caption{\small \label{fig:t_H} Spectra of cosmic rays and neutrinos from cosmic-ray reservoirs with black hole jets. Same as in the fiducial mixed-composition scenario but with the cosmic-ray injection time $t_{\rm inj}=t_H$. 
}
\end{figure*}

Our simulations allow us to approximately calculate spectra of neutrinos and cosmic rays for a steady injection over a duration $t_{\rm inj}\sim1-10$~Gyr. The history of injections from AGN is time-dependent and is dominated by their past activities at $z\sim1-2$. Considering this effect, we have adopted $t_{\rm inj}=2$~Gyr. 
Two effects induce a spectral steepening in our model. First, a spectral break due to the escape of cosmic rays occurs at sufficiently high energies when $t_{\rm diff}(E)<t_{\rm inj}$~\cite{Murase:2008yt,Murase:2013rfa}. In addition, a break in the injection spectrum is caused by the confinement of cosmic rays in the cocoon. Figure~\ref{fig:t_H} demonstrates that $t_{\rm inj}=t_H$, where $t_H=13.75~{\rm Gyr}$ is the age of the Universe, leads to similar results. Thus for our fiducial parameters, our results are insensitive to changes in $t_{\rm inj}$ which affects the confinement ability of the cluster.
In essence, since we achieve $s<s_{\rm acc}$ (where $s$ is the effective injection spectral index), our model is consistent with model-independent bounds on cosmic-ray reservoir scenarios, in which the effective injection spectral index, $s$, is constrained by the $\gamma$-ray background to be smaller than $s\sim2.1-2.2$~\cite{Murase:2013rfa}. 

All cosmic-ray ions and neutrinos, including the primary cosmic-ray particles injected into the simulation and their secondary and higher-order products, are tracked down to 1 TeV. The neutrino spectrum below this energy is extrapolated based on a spline fit to the simulation results between 2.2 TeV and 10 TeV. From the neutrino spectrum we obtain the spectrum of their $\gamma$-ray counterparts by~\cite{Murase:2015xka}
\begin{equation}\label{eq:Qgamma}
E_{\gamma}Q_{E_\gamma}\approx\frac{2}{3}(E_{\nu}Q_{E_\nu})\big|_{E_\nu=E_\gamma/2}\,,
\end{equation}
where $E Q_{E}$ is the energy generation rate density per logarithmic energy, and $\gamma$-ray and neutrino energies are related as $E_\gamma\approx2E_\nu$. Then we take into account electromagnetic cascades during the intergalactic propagation by solving transport equations~\cite{Murase:2012df}. Note that clusters and groups are expected to be transparent up to $\sim100$~TeV energies~\cite{Inoue:2005vz,Murase:2012rd}. Either leptonic or leptohadronic, $\gamma$-ray emission can be observed directly from jets and lobes of radio-loud AGNs.  In particular, the emission from blazars jets significantly contributes to the EGB. Primary relativistic electrons cool inside the AGN, and are not expected to diffusively enter the ICM like cosmic-ray ions.

The diffuse flux computation convolves the mass function of reservoirs (that are clusters and groups) and luminosity density of all possible cosmic-ray accelerators (that are mainly AGN jets) as shown by equation~1. Since it is assumed that all radio-loud AGNs, including BL Lac objects and their misaligned counterparts, can produce UHECRs (see reference~\cite{Murase:2012rd} for the discussion on the inner jet model for UHECRs), we use a luminosity density obtained by integrating the luminosity function. Note that the photon luminosity densities of radio galaxies, blazars, and AGN (rather than the number densities) show a positive redshift evolution~\cite{Willott:2000dh,Ajello:2013lka,Ueda:2014tma}, and uncertainty in the redshift evolution can affect the diffuse fluxes only by a factor of $\sim3$ (see reference~\cite{2016PhRvD..94j3006M} for more details).
Then, we simulate host clusters in 10 mass bins from $M_{\rm min}=5\times10^{13}\,M_\odot$ to $M_{\rm max} = 10^{16}\,M_\odot$, and 14 redshift bins from $z_{\rm min} =0.005$ to $z_{\rm max} =5$. The result, however, barely depends on $M_{\rm max}$ and $z_{\rm max}$ since the population of massive clusters at high redshifts is negligible. The minimum redshift $z_{\rm min} = 0.005$ corresponds to the distance of the Virgo cluster at $\sim 20$~Mpc. The halo mass function is calculated by 
\begin{equation}
\label{eq:dndM}
\frac{dn}{dM} (M, z)= f(\sigma) \frac{\rho_m}{M}\frac{d \ln~\sigma^{-1}}{dM}
\end{equation} 
where $\rho_m(z) =\rho_m(0)\,(1+z)^3$ is the mean density of the universe at given redshift, and $\sigma(M, z)$ corresponds to the variance of the linear density field smoothed on a top-hat window function~\cite{2008ApJ...688..709T} $r_t=(3M/4\pi\rho_m)^{1/3}$. For $f(\sigma)$, we adopt the Sheth-Tormen mass function multiplicity~\cite{sheth_tormen_mass_function}, which is consistent with results from N-body cosmological simulations~\cite{Jenkins01,Warren:2005ey,2008ApJ...688..709T}. 

A galaxy cluster and group naturally hosts supermassive black holes with powerful jets as cosmic-ray accelerators, which may include not only radio-loud AGN but also radio-quiet AGN~\cite{Murase:2015ndr}. The emission from the accelerators themselves is a  very different subject beyond the scope of this work. As noted above, synchrotron emission from primary electrons in the jets (or lobes) of blazars and radio galaxies have been observed. Since the injected cosmic-ray ion flux may be $\sim100$ times higher than the electron flux, as indicated by the Galactic cosmic-ray data, it is also possible to have a hadronic component as often discussed in the leptohadronic model for blazars. Our model is consistent with either leptonic or leptohadronic model for the high-energy emission from the cosmic-ray accelerators, and the blazars may make a subdominant contribution to the PeV neutrino flux. This is a generic feature of cosmic-ray reservoir scenarios.
In principle, cosmic-ray accelerators may even be powerful transients in galaxies including $\gamma$-ray bursts, supernova explosions, fast-rotating pulsars, tidal disruption events~\cite{Waxman:2013zda,Senno:2015tra}, and blazar flares~\cite{Dermer:2014vaa}. In addition, particle acceleration can also happen in the accretion shocks that are commonly found at the outskirts of large-scale structures~\cite{Kang:1995xw,Inoue:2007kn}. 
%Cosmic rays accelerated by accretion shocks can contribute to the second knee, and neutrinos have been suggested as a probe of such a scenario~\cite{Murase:2008yt}. 
The accretion shock contribution is expected to be more important for massive clusters~\cite{Murase:2008yt,FO16,2016PhRvD..94j3006M}, and the corresponding neutrino production has been subject to various constraints~\cite{Kushnir:2009vp,FO16,2015A&A...578A..32Z,2016PhRvD..94j3006M}. In contrast, the contribution from internal accelerators such as AGN can be more important for less massive clusters~\cite{Murase:2008yt,Kotera:2009ms,Murase:2013rfa}, since the population of such clusters is higher and AGN are found to have a strong redshift evolution. The distribution of higher-energy cosmic rays is expected to be more uniform~\cite{Keshet:2010aq}.

For the extragalactic propagation, we use CRPropa 3 and take into account the photomeson and photodisintegration interactions between cosmic rays and background photons including the CMB and the EBL (where the best-fit model of reference~\cite{2004A&A...413..807K} is used), as well as the decay of unstable intermediate products and the Bethe-Heitler production of electron-positron pairs. We use the transmission factor~\cite{Kotera08a} to take into account the magnetic horizon effect due to structured EGMFs~\cite{Lemoine:2004uw,2008ApJ...682...29D, 2014JCAP...11..031A}. The confinement in the structured EGMFs may also lead to additional secondary production especially below the ankle, which is not included to save computation time.
Cosmogenic neutrinos~\cite{Beresinsky:1969qj} are generated by photomeson interactions of protons above the Greisen-Zatsepin-Kuzmin energy~\cite{1966PhRvL..16..748G,1966JETPL...4...78Z} ($E_{\rm GZK}\sim6\times10^{19}$~eV) during their intergalactic propagation. The flux of cosmogenic neutrinos is lower if UHECRs are dominated by intermediate or heavy nuclei, as opposed to pure protons~\cite{2010PhRvD..81l3001M,KO11,Ahlers:2012rz}. Cosmogenic $\gamma$ rays from intermediate and heavy nuclei are dominated by the Bethe-Heitler pair production~\cite{Ahlers:2011sd,Decerprit:2011qe,2012ApJ...749...63M}. It is known that cascaded $\gamma$ rays follow a universal spectral shape, and that their flux is basically determined by the energy injection rate of cosmic rays at ultrahigh energies~\cite{Berezinsky:1975zz,Wang:2011qc,Murase:2012df}.  For demonstration purposes, the cluster contribution to the EGB is estimated by adding the cosmogenic $\gamma$-ray flux of a mixed composition scenario~\cite{Decerprit:2011qe} to the source $\gamma$-ray flux obtained in this work. As in the simplest convergence theory~\cite{2016PhRvD..94j3006M}, the total $\gamma$-ray flux from our model is compatible to the non-blazar component of the EGB~\cite{PhysRevLett.116.151105,Lisanti:2016jub}. 

The combined fit to the Auger spectral data, the $\langle X_{\rm max}\rangle$ data, and the IceCube spectral data is performed as follows. The goodness-of-fit is assessed with a total $\chi^2$ contributed by three separate fits, given that the UHECR energy spectrum, the $X_{\rm max}$ distributions, and the neutrino energy spectrum are independent measurements:
\begin{eqnarray}
\chi^2 (C_{\rm norm}, \delta_{E}^{\rm CR}, \delta_{E}^{\nu}) = \chi^2_{\rm CR,spec}(C_{\rm norm}, \delta_{E}^{\rm CR}) + \chi^2_{{\rm CR}, X_{\rm max}} + \chi^2_{\nu,\rm spec}(C_{\rm norm}, \delta_E^\nu).
\end{eqnarray}
$C_{\rm norm}$ is a universal factor that normalizes the predicted fluxes of UHECRs and neutrinos,  $\delta_{E}^{\rm CR}$ and $\delta_{E}^{\nu}$ denote the energy-scale displacements in the UHECR and neutrino spectra, defined so that $E\rightarrow E'=(1+\delta_E)\,E$. The allowed displacements are capped by the systematic energy uncertainties in the Auger spectrum, $\Delta_E^{\rm CR} = \pm14\%$~\cite{Aab:2015bza}, and the deposited energy resolution of cascade events of IceCube~\cite{IceCube_ICRC}, $\Delta_E^\nu \sim \pm15\%$. We do not allow an energy-scale displacement in the fits to the $\langle X_{\rm max}\rangle$ data since the energy resolution of the Fluorescence Detector (FD) of Auger is narrower than the width of the energy bins used here~\cite{2016arXiv161207155T}. Specifically, the $\chi^2$ functions are defined as (see e.g., reference~\cite{2015APh....62...66B}):
\begin{eqnarray}
\chi_{\rm CR/\nu,\,spec}^2&=&\sum_i\left(\frac{\Phi_{\rm CR/\nu} (E_i',C_{\rm norm})-\left(\Phi_{\rm CR/\nu}\right)^{\rm obs}_i}{{\left[\Delta\left(\Phi_{\rm CR/\nu}\right)\right]}^{\rm obs}_i}\right)^2 + \left(\frac{\delta_E^{\rm CR/\nu}}{\Delta_E^{\rm CR/\nu}}\right)^2, \\ \nonumber
\chi_{{\rm CR},X_{\rm max}}^2&=&\sum_{j}\left(\frac{\langle X_{\rm max}\rangle(E_j) - \langle X_{\rm max}\rangle^{\rm Auger}_j} {{\left[\Delta\langle  X_{\rm max}\rangle\right]}^{\rm Auger}_j }\right)^2. 
\end{eqnarray}
Here $\Phi_{\rm CR}$ and  $\Phi_\nu$ are the flux of UHECRs measured by the Auger Surface Detector (SD)~\cite{Aab:2015bza} and the flux of astrophysical neutrinos in the six-year IceCube data~\cite{Aartsen:2016xlq2}, $\langle X_{\rm max}\rangle$ is the mean of $X_{\rm max}$ measured by the Auger FD~\cite{2014PhRvD..90l2005A, Aab:2015bza}, and $\Delta \left(\Phi_{\rm CR}\right)$, $\Delta \left(\Phi_\nu\right)$, and $\Delta \langle X_{\rm max}\rangle$ are the corresponding uncertainties defined via the quadrature sum of the statistical and systematic uncertainties of the measurements. We use $E^3\Phi_{\rm CR}$ and $E^2\Phi_\nu$ to evaluate the fits to cosmic-ray and neutrino data respectively as these are the forms of data provided by the corresponding experiments. The systematic uncertainty of the UHECR spectrum is derived from that of vertical Auger SD data sets recorded by the 750 m and 1500 m arrays, as well as inclined events recorded by the 1500 m array~\cite{Aab:2015bza}, weighted by the event number from each data set in each energy bin. In addition, as suggested by a comparison of the TA and Auger spectra~\cite{2017arXiv170509111V}, we increase the energies of Auger spectrum by 5\% (so that $-19\% <\delta_E^{\rm CR} < 9\%$) to determine the absolute flux of UHECRs. The data used in the fit consists of the Auger spectrum in 18 bins with 0.1 increment of $\log_{10}(E/\rm eV)$ from 18.45 to 20.15, $\langle X_{\rm max}\rangle$ in 12 bins with 0.1 increment from 18.45 to 19.50, and one last bin between 19.50 and 20.00, and the IceCube spectrum of high-energy starting events above $2\times{10}^{14}$~eV. A total of 33 non-zero data points are used for the fit. When the upper and lower errors of a data point are different, the upper error is used when the model prediction is higher than the data, and vice versa. 

The default parameters are set to be $E_{\rm max} = 2\times10^{21}/Z$~eV and $s_{\rm acc}=2.3$ as motivated by black hole jets of radio-loud and radio-quite AGN~\cite{2012ApJ...749...63M,Peer:2009vnw}. Our results are however not very sensitive to small variations of these parameters. Generally a softer cosmic-ray spectrum also leads to a softer neutrino spectrum, and a larger $E_{\rm max}$ leads to higher flux of cosmogenic neutrinos.  
We inject a chemical composition based on the abundance of elements in Galactic cosmic rays. Specifically, we adopt the elemental abundances at PeV$/Z$ based on cosmic-ray measurements from 10~GeV to 100~PeV~\cite{2003APh....19..193H}. The iron abundance is enhanced by a factor of 2, since elliptical galaxies may have higher metallicity than star-forming galaxies such as the Milky Way. The energy flux ($E^2dN_{\rm inj}/dE$) at the same rigidity scales to (0.625 H, 0.252 He, 0.053 CNO, 0.009 Si, 0.124 Fe), which is also compatible with the cosmic-ray composition at 100~PeV. 
By minimizing $\chi^2$ through the fits with the Auger and the IceCube data, we find a best-fit $\chi^2_{\rm dof}=44.5/ 30=1.48$ (corresponding to a non-chance occurrence probability of 4.3\% for the fiducial case to explain the data) at $\hat{\delta}_E^{\rm CR}=-0.12$ and $\hat{\delta}_E^\nu=-0.01$. The best-fit $C_{\rm norm}$ corresponds to $f_{\rm jet}=23.7\%$, assuming $L_{\rm CR} = 10^{45}\,\rm erg\,{\rm s}^{-1}$ above 10~GeV. We obtain similar results, $\chi^2_{\rm dof}\sim1.5-2$, if changing the starting energies to $10^{18.45\pm0.2}$~eV in the UHECR fits and $6\times{10}^{13}$~eV in the neutrino fit. The fit could be improved to $\chi^2_{\rm dof}=1.3$ with the non-blazar component of the Fermi data above 50~GeV, although more careful analyses are required to properly estimate the uncertainties of the $\gamma$-ray data. A fit with $\chi^2_{\rm dof} \sim 3-5$ is obtained without applying the EGMF. 
 A more dedicated fitting allowing arbitrary abundance, $E_{\rm max}$, $s_{\rm acc}$, and varying EGMF parameters can potentially provide a better fit to data~\cite{2016arXiv161207155T, Wittkowski:2017okb}, but such a study is beyond the scope of this work. Given that there are uncertainties in the acceleration and propagation models as well as photonuclear interactions, our work is sufficient for demonstrating that the physical connection among multi-messengers is achieved by a concrete model with the physically-motivated source parameters.

The connection between PeV neutrinos and UHECRs in cosmic-ray reservoir scenarios was suggested before the discovery of IceCube neutrinos~\cite{Murase:2008yt,Kotera:2009ms}, and their link to the diffuse isotropic $\gamma$-ray background has also been discussed in references~\cite{Murase:2013rfa,2016PhRvD..94j3006M,PhysRevD.95.063007}. 
Our model is the first concrete model that simultaneously explains the measured diffuse fluxes of all three messengers without violating observational constraints. The model avoids {\it ad hoc} tuning of parameters, in the sense that the source population (including the number density and redshift evolution), the profiles of target gas and radiation, and the injection composition are all determined by observations, and the choice of free parameters ($E_{\rm max}$ and $s_{\rm acc}$) is physically motivated. Our numerical approach takes into account impacts of magnetic structures within sources and contributions from secondary particles. 
Note that non-flaring blazar models~\cite{2014PhRvD..90b3007M,PhysRevD.92.083016} as the dominant sources of IceCube neutrinos are disfavored by the stacking analyses~\cite{Aartsen:2016lir} and multiplet analyses~\cite{2016PhRvD..94j3006M} of the IceCube data. High-redshift sources or rapidly evolving sources such as quasars~\cite{PhysRevD.95.063007,Xiao:2016rvd} do not explain the Fermi data above 10 GeV due to the strong EBL absorption.
Unlike the blazar and high-redshift source models, our results respect the updated constraints on the non-blazar component of the Fermi EGB~\cite{PhysRevLett.116.151105} with the important contribution from cosmogenic $\gamma$ rays~\cite{2016PhRvD..94j3006M}, the neutrino flux above 10~PeV~\cite{2016arXiv160705886I} by including cosmogenic neutrinos, and the heavy-rich composition of UHECRs indicated by Auger~\cite{Aab:2015bza}, as well as the sub-ankle component above ${10}^{17}$~eV.

\section*{Light composition scenario}
\begin{figure}[tb]
\centering\includegraphics[width=0.60\linewidth]{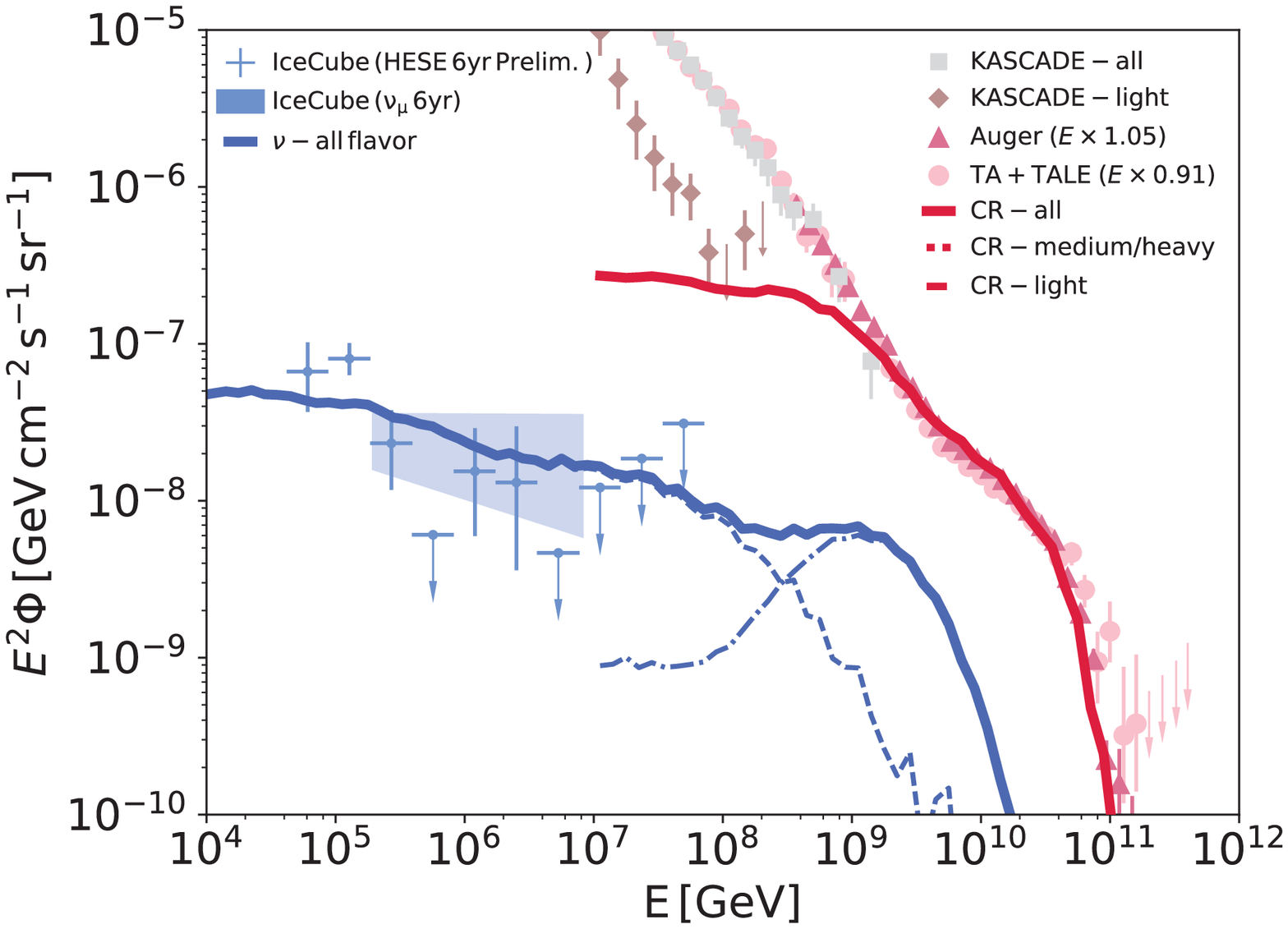}
\caption{\small \label{fig:alt} Spectra of cosmic rays and neutrinos from cosmic-ray reservoirs with black hole jets, assuming that  all cosmic rays are protons with $E_{\rm max}=10^{19.6}$~eV. The propagated cosmic-ray spectrum (solid red) matches the Auger and TA data above the ankle ($\sim 10^{18.5}$~eV).
}
\end{figure} 

So far we have mainly focused on a mixed composition scenario, but the interpretation of the chemical composition of UHECRs is under debate. While three interaction models, EPOS~\cite{2013JCAP...02..026P}, QGSJET~\cite{Ostapchenko:2006wc} and SIBYLL~\cite{Ahn:2009wx} are commonly used, hadronic interaction models at the highest energies are based on the extrapolation from experimental data at lower energies, and thus still suffer from non-negligible uncertainties. 
A simultaneous fit to the Auger spectrum, $\langle X_{\rm max}\rangle$ and $\sigma(X_{\rm max})$ suggests a dominance of intermediate or heavier nuclei at the injection site~\cite{Taylor:2011ta,2016arXiv161207155T}, which is quite challenging for known astrophysical sources. References~\cite{2016arXiv161207155T, Wittkowski:2017okb} find a main minimum and a second minimum in their fitting. The main minimum requires a hard injection spectral index $s=0.9$ (or 1.61 with the EGMF), which is different from typical expectations from the diffusive shock acceleration mechanism (although such a hard spectrum could be reached by the shock acceleration at ultrarelativistic shocks~\cite{2008A&A...492..323M,Aoi:2007aj} or other mechanisms such as the shear acceleration~\cite{Kimura:2017ubz}). The second minimum suggests $s=2.0$ (or 2.3 with the EGMF) and is similar to the mixed composition scenario shown here, but with a $\sigma(X_{\rm max})$ lighter than the Auger measurements between $10^{18.8}$~eV and $10^{19.4}$~eV. The UHECR composition can also be modified by interactions inside cosmic-ray accelerators~\cite{2012ApJ...749...63M, 2012ApJ...750..118F, 2014PhRvD..90b3007M}. More quantitative discussion on the composition is beyond the scope of this work, and we do not attempt to fit to the $\sigma (X_{\rm max})$ data. 

Below we consider an alternative scenario, where cosmic rays from radio-loud AGN are dominated by pure protons ({\it aka} the ``proton scenario"). Such a possibility is shown in Figure~\ref{fig:alt} for the completeness of our demonstration, since it has been argued that $X_{\rm max}$ measurements by TA may be consistent with the light composition~\cite{TA_ICRC15}. Except for a different composition, we adopt the same injection spectrum, $dN_{\rm inj}/dE\propto E^{-2.3}$ and the same maximum energy, $E_{\rm max}=10^{19.6}$~eV, as in our mixed composition case. 
Like in the mixed-composition scenario, the neutrino products from the ICM is comparable to the $0.1-1$~PeV measurements by IceCube. The difference between a nucleus and a proton in neutrino production can be very roughly estimated as follows. Considering that the inelastic cross section of a nucleus with mass number $A$ scales as $\sigma_{Np}\sim A^{2/3}\,\sigma_{pp}$, and that a nucleus with $Z\times E$ experiences the same level of diffusion as a proton with $E$, the interaction between the nucleus and the ICM gas has an effective optical depth $f_{Np}(ZE)\sim(\kappa_{Np}/\kappa_{pp})A^{2/3}\,f_{pp}(E)$, where $\kappa_{Np}$ and $\kappa_{pp}$ are inelasticities of the projectile particle in each case. Therefore, a neutrino product from a nucleus and that from a proton with $Z$ times lower energy have an energy ratio $Z/A\sim0.5$ and an energy flux ratio $\sim(\kappa_{Np}/\kappa_{pp})A^{2/3}$.  

%One feature of the proton scenario is the presence of a prominent EeV neutrino bump due to a copious production of cosmogenic neutrinos. The energy flux level of neutrinos is still consistent with the current upper limits posed by the non-detection of extremely high-energy neutrinos in the IceCube data~\cite{2016arXiv160705886I}, but should be detectable in the near future.  In the classical ankle scenario with $s=2.0$, the diffuse isotropic $\gamma$-ray flux is shown to be comparable to the non-blazar component of the EGB~\cite{2016PhRvD..94j3006M}. Our proton scenario with $s_{\rm acc}=2.3$ effectively leads to $s\sim2$, but the tension with the $\gamma$-ray data is slightly stronger due to more cosmogenic $\gamma$-ray production.

\end{document}